\documentclass[]{aa}
\usepackage[dvips]{graphicx}
\usepackage{amssymb}
\newcommand{\mcol}[3]{\multicolumn{#1}{#2}{#3}}

\newcommand{\Msol}{\mbox{\rm M$_{\odot}$}}

\begin{document}
\thesaurus{
	    (08.16.4;  
	    11.19.4;  
	    11.19.5;  
	    13.09.1)}  

\title{How to search for AGB stars in near-IR post-starburst spectra}


\author{A.\,Lan\c{c}on\inst{1}, M.\,Mouhcine\inst{1}, 
M.\,Fioc\inst{2}\fnmsep\thanks{Now at: NASA/GSFC, Greenbelt, MD 20771, USA}, 
D.\,Silva\inst{3}}

\offprints{A. Lan\c{c}on}

\institute{Observatoire Astronomique, Universit\'e L.\,Pasteur \& CNRS: 
UMR 7550, 11 rue de l'Universit\'e, F-67000 Strasbourg
\and
Institut d'Astrophysique de Paris, 98\,bis Bd. Arago, F-75014 Paris
\and
European Space Observatory, Karl-Schwarzschild-Str.2, D-85748 Garching}

\date{Received ; accepted ..}

\authorrunning{Lan\c{c}on et al.}
\titlerunning{Near-IR AGB signatures in stellar populations}
\maketitle

\begin{abstract}
Based on evolutionary spectral synthesis models explicitly including
the spectra of variable AGB stars, we select near-IR
features that dentify the strong O-rich or C-rich AGB contributions 
to the near-IR light of post-starburst populations. AGB temperature
scales and lifetimes remain major sources of uncertainties. We discuss
applications and suggest massive post-starburst clusters as prime
targets for observational tests.

\keywords{stars: AGB -- galaxies: star clusters --
galaxies: stellar content -- infrared: galaxies}

\end{abstract}

\vspace*{-0.4cm}
\section{Introduction}
Population synthesis and star counts in clusters indicate that
asymptotic giant branch stars (AGB stars) 
contribute more than 50\,\% of the integrated $K$ band light of stellar 
populations with ages of a few $10^8$\,yrs, i.e.
after the red supergiants have died and before the less luminous
but more numerous first giant branch stars (FGB) have taken over
(Bruzual \& Charlot 1993, Ferraro et al. 1995). 
But how can one recognise AGB contributions in integrated galaxy
light? While star forming regions are indicated by line emission and 
older starbursts by the strong CO absorption of red supergiants 
(e.g. at 2.3\,$\mu$m), it remains a challenge to separate AGB and FGB 
contributions in more evolved near-IR spectra. 

The AGB extends to higher luminosities and lower effective temperatures
(T$_{\rm eff}$)
than the FGB but the consequences on the surface brightness and near-IR
colours of integrated populations are difficult to disentangle
from those of inhomogeneous extinction or of the metallicity dependence
of AGB evolution tracks. The way to proceed is to exploit the
spectral signatures resulting from Mira-type pulsation in the later AGB
stages.

Although Johnson \& M\'endez (1970) already observed that variable
AGB stars (LPVs) display much stronger
near-IR H$_2$O absorption bands than static giants or supergiant stars,
subsequent spectrophotometric studies of cool stellar populations
did not attempt to discuss LPVs and static giants separately.
Indeed, the separation
was omitted until now in the stellar libraries used as input to population
synthesis codes (Lan\c{c}on \& Rocca-Volmerange 1992,
Terndrup et al. 1991, Lejeune et al. 1998). 
Model atmospheres confirm that the deep molecular features observed in LPVs
are intimately linked to Mira-type pulsation (Bessell et al. 1989a, 1996), 
but the trends are not reproduced quantitatively (Mouhcine \& Lan\c{c}on 1998)
and the model spectra are not reliable enough to be included in the libraries. 
The present study is  therefore based on new observational spectroscopic data.

After a brief description of the updated population synthesis tool, 
we select near-IR narrow-band filter diagnostic plots for searches of 
AGB-dominated populations.  Depending on
the available instrumentation, the photometry can be replaced profitably 
by low resolution spectroscopy. We then discuss some of the now
feasible and most promising applications.

\section{Population synthesis models including LPV spectra}

\subsection{New input spectra for cool stars}
\label{data.sect}

The data set (Lan\c{c}on \& Wood 1997) 
consists of 140 me\-dium resolution ($R=1100$)
near-IR spectra obtained with the cross-dispersed grisms of the 
camera {\sc caspir} at the 2.3m
ANU Telescope at Siding Spring (McGregor 1994) 
and more than 150 low resolution optical spectra 
(Rey\-nolds Spectrograph, Mt Stromlo Observatory 1.9m Telescope).
About 100
pairs, i.e. quasi-simul\-taneous observations of the 
same star in the two wavelength ranges, provide full
wavelength coverage from 0.5 to 2.5\,$\mu$m for
a sample of LPVs. Carbon stars, static
giants and cool supergiants are included in the set.

\begin{figure}
\vspace*{-0.3cm}
\includegraphics*[clip=,angle=0,width=9cm,height=8cm]{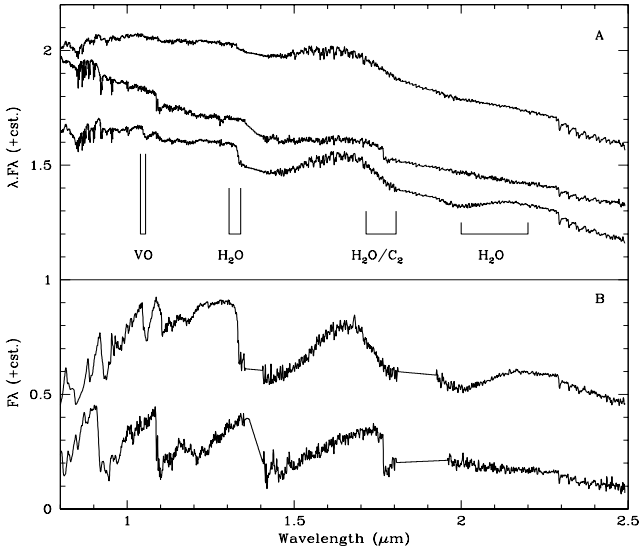}
\caption[]{{\bf A:}
Synthetic post-starburst spectra (instantaneous burst, Z=0.008).
Bottom: t=200\,Myr, LPV atmospheres
remain O-rich ($\lambda F_{\lambda}+1$); middle: t=200\,Myr, but
the dominant LPVs have become C-rich ($\lambda F_{\lambda}+1.15$); top:
giant dominated population (3\,Gyr). Band
centers/widths in nm are 1039.5/5, 1054/6, 1305/20, 1340/16,
1715/70, 1805/70, 2000/80, 2200/110.
{\bf B:} The cool mira R Phe (top)
and the C star S Cen (bottom), both observed in Jan. 1996.}
\label{filters.fig}
\end{figure}

The data confirm that Miras produce weaker CO features than red supergiants
but can display deeper and broader H$_2$O bands than any static star, 
even when their energy distribution indicates
relatively warm T$_{\rm eff}$ values ($\geq$3000\,K).
The coolest LPVs also display near-IR VO and TiO absorption
bands between 1 and 1.3\,$\mu$m. Carbon star spectra are dominated
by CN absorption, and easily recognised by the sharp C$_2$
bandhead at 1.77\,$\mu$m (Fig.\,\ref{filters.fig}B).

Temperatures were assigned to instantaneous stellar spectra
in two ways considered as extremes, and an intermediate scale
was adopted in the synthesis calculations (unless otherwise stated). 
The first (and higher) scale results from the comparison with
the static giant model atmospheres described in Alvarez \& Plez (1998),
the second from the angular diameter measurements compiled by Feast (1996).
The uncertainties in LPV temperature scales are a fundamental issue that will 
be discussed extensively elsewhere (Mouhcine et al., in preparation).

\subsection{Modelling the spectral evolution of stellar populations}
The evolution of synthetic stellar populations in the HR diagram
is computed with the population synthesis code {\sc P\'egase}
(Fioc \& Rocca-Volmerange 1997) and its extension to
non-solar metallicities (Fioc 1997).
Although the code is able to compute and follow the continuous chemical 
evolution of a galaxy, we restrict ourselves here to constant,
solar ($Z=0.02$) or Large Magellanic Cloud ($Z=0.008$) metallicities
in order to avoid additional evolution parameters and to work with
AGB evolution models that have been reasonably well tested against
observations. 
A Salpeter IMF is assumed (power law index -2.35, lower
mass limit 0.1\,\Msol).

The stars evolve up to the early AGB along the tracks of
Bressan et al. (1993) and Fagotto et al. (1994). The extensions of Fioc (1997)
through the thermally pulsing AGB phase (TP-AGB)
follow the prescriptions of Groenewegen et al. (1993, 1995) with only
a slight adjustment in temperature (well inside theoretical uncertainties).
In the HR diagram, the resulting tracks superimpose extremely well on those
of Vassiliadis \& Wood (1993). 

The new spectroscopic data are used in the cool, luminous regions
of the HR diagram, as an addition to the colour-corrected spectral
library of Lejeune et al. (1998). 
The LPV spectra are averaged in 0.025 wide logarithmic
temperature bins (Lan\c{c}on 1998). The resulting
sequence displays a regular evolution with T$_{\rm eff}$,
both in colours and molecular band absorption indices. 

\section{Near-IR indices to search for AGB stars}

Figure\,\ref{filters.fig} defines a set of filters well suited for measurements
of the specific LPV features described in Sect.\,\ref{data.sect}.
For practical (observational) reasons, the passbands
are chosen among those of existing filter sets whenever possible
(``Wing filters", Bessell et al. 1989b; Hubble Space Telescope NICMOS filters).
All indices are flux ratios expressed in magnitudes  
and take the value 0 for Vega.

\begin{figure}
\vspace*{0.1cm}
\includegraphics*[clip=,angle=270,width=8.8cm]{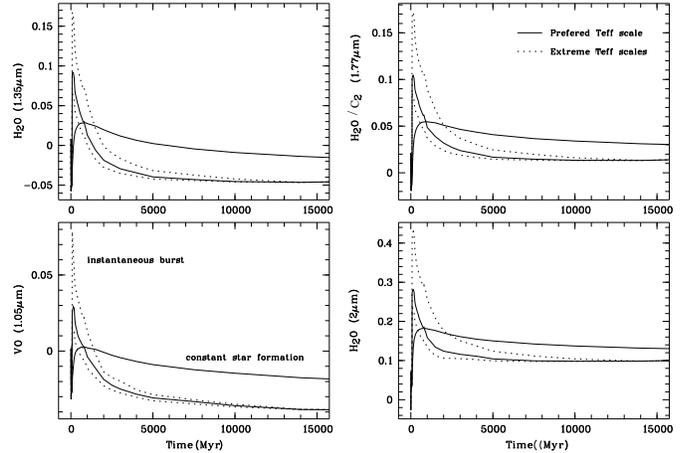}
\caption[]{The time evolution of selected indices.  
Measuring changes of a few percent in indices close to the variable
telluric H$_2$O absorption bands (cut out in Fig.\,1\,B) 
requires a dry observing site
and particularly frequent observations of a reference star close to
the target population.
}
\label{evol.fig}
\end{figure}

The time evolution of the selected molecular indices at solar metallicity
is shown in Fig.\,\ref{evol.fig} for an instantaneous burst and for constant
star formation, with the assumption that all LPVs are oxygen rich.
Both H$_2$O and VO indices clearly identify
the post-starburst period, i.e. populations with ages between
$10^8$ and $10^9$\,yrs.

The most sensitive index is the measure of H$_2$O at 2\,$\mu$m with
respect to the $K$ band ``continuum";
it displays the largest variations with time.
However, this index would fail to identify the C-rich LPVs, which
may represent a significant fraction of the luminous AGB 
stars depending on the environment (Habing 1996). We therefore 
consider it essential to include the  1.77\,$\mu$m index in the observational
campaigns: while it detects H$_2$O absorption in O-rich Miras, it has
the useful property to also detect the C$_2$ bandhead of C-rich LPVs.

\begin{figure}
\includegraphics[clip=,angle=270,width=9cm]{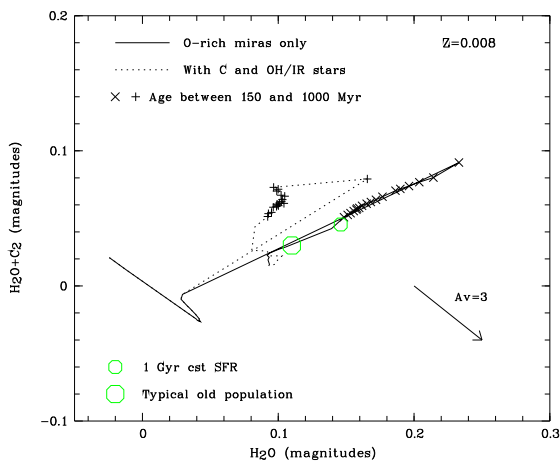}
\caption[]{Synthetic evolution in the prefered two-index diagram. The
``jump" to high index values around 100\,Myr takes less than 50\,Myr.
Circles identify plausible locations of contaminating older populations.}
\label{models.idx.fig}
\end{figure}

We have computed a sequence of models in which the LPVs
become carbon rich after the fraction of the TP-AGB lifetime
indicated by Groenewegen \& de Jong (1993, 1994).  
Results are presented in the diagnostic plot in Fig.\,\ref{models.idx.fig}.
The results are qualitatively independent 
of the prescription a\-dopted for the O to C transition, and of the
detailed filter passbands. A high 1.77\,$\mu$m H$_2$O/C$_2$ index unambiguously
identifies the predominance of AGB stars and thus post-starburst populations:
there is no confusion with younger or older populations. The second index  
then indicates the chemical nature of the dominant AGB stars once
they have been found. 

\begin{table*}
\caption[]{Minimum contribution of the post-starburst allowing its
unambiguous detection, for 2 extreme representations
of the underlying population and 4 detection criteria (solar metallicity,
Salpeter IMF, lower mass=0.1\,\Msol, O-rich AGB).}
\label{dil.tab}
\begin{tabular}{|c||c|c|c|c||c|c|c|c|} \hline
          & \mcol{4}{|c|}{Old 1} & \mcol{4}{|c|}{Old 2} \\ \hline
Selection threshold & Mass ($\%$) & $L_V (\%)$ & $L_I (\%)$ & $L_K (\%)$ &
            Mass ($\%$) & $L_V (\%)$ & $L_I (\%)$ & $L_K (\%)$ \\ \hline
I$_{H_2O(2\mu m)}=0.19$ & {\bf 9} & 72 & 57 & 52 & {\bf 11} & 18 & 17 & 19 \\
I$_{VO(1.05\mu m)}=0.003$ & {\bf 11} & 78 & 65 & 60 & {\bf 7} & 13 & 12 & 14 \\
I$_{H_2O+C_2(1.77\mu m)}=0.056$ & {\bf 8} & 70 & 55 & 50 & {\bf 5} & 10 & 9 & 10 \\
I$_{H_2O(1.35\mu m)}=0.03$ & {\bf 10} & 76 & 62 & 57 & {\bf 8.5} & 16 & 14 & 16 \\ \hline
\end{tabular}
\vspace*{-8pt}
\end{table*}

\section{Discussion and applications}

\label{disc.sect}

The numerical model predictions are limited by the uncertainties
in the TP-AGB evolution models and in the assignment of a representative
spectrum to each point of the theoretical HR diagram. TP-AGB lifetimes
vary by a factor of two from one author to the other 
(Vassiliadis \& Wood 1993, Groenewegen et al. 1995, Marigo et al. 1998).
Those implemented in {\sc P\'egase} are among the shortest ones:
it is unlikely that the TP-AGB contributions be overestimated.

The dotted lines in Fig\,\ref{evol.fig} demonstrate the sensitivity of the 
post-starburst AGB signatures to the combined 
temperature scales of the spectra and the evolutionary tracks. The
situation is similar to that faced in attempts to determine
red supergiant contributions from 2.3\,$\mu$m CO absorption: very
strong features carry a clear message, but more common lower values
are not conclusive. It will be
essential to test the models with simple stellar populations
of known ages, before attempting to infer quantitative
AGB contributions from the integrated spectra of more complex objects.
\medskip

How robust are the molecular post-starburst features with respect
to dilution in the light of an underlying, more evolved population?

Results are given in Tab.\,\ref{dil.tab}. Without a priori 
know\-ledge of the history of the underlying population (but assuming that
it is older than $\sim \!\!2$\,Gyr and obeys the same stellar evolution
prescription as the postburst), the unambiguous detection of a postburst 
via its AGB signatures requires that the observed molecular indices
exceed those constantly star-forming populations can reach
(cf. Fig.\,\ref{evol.fig}). We may consider two extreme underlying
populations.
``Old 1" is a 10\,Gyr old instantaneous burst remnant. Its near-IR
spectrum is dominated by relatively early type red giants
with H$_2$O or VO features far below the adopted postburst detection
threshold; in this case, exceeding the threshold
requires a large contribution of the postburst to the integrated light. 
``Old 2" is the result of 2\,Gyrs of  constant star formation.
Its spectrum is still contaminated by AGB stars, 
and its molecular bands are just below the adopted threshold themselves; 
hence, integrated indices exceeding the threshold may
correspond to a much smaller contribution of the  postburst to the light.

Incidentally, the mass-to-light ratios of the underlying populations compensate
in such a way that a postburst detection based on the above
criterion will, in both cases, imply that the postburst represents
more than $\sim 10\,\%$ of the mass in the observed field of view.
\medskip

Another hurdle to keep in mind is that TP-AGB stars, while intrinsically
bright, are rare objects. At ages of a few 10$^8$\,yrs,
the stochastic fluctuations in the near-IR
fluxes due to TP-AGB number fluctuations are significant 
($10\,\%$ or more)
in postburst populations containing less than about $10^6\,\Msol$\ of
stars (Ferraro et al. 1996, Santos \& Frogel 1997, Lan\c{c}on 1998).

In summary, the ideal target populations should have ages between 
$10^8$ and $10^9$ years, and should be both massive and not too
heavily diluted.  Dilution can be either intrinsic, if young and old stars
are mixed efficiently (e.g. in E+A galaxies),
or observational, if the spatial  resolution of the data is 
poor. We will not discuss nearby objects here, for which
high resolution imaging nowadays directly provides very complete information.
Interesting spectrophotometric
applications are found in their more distant counterparts or progenitors.
\medskip
 


Many observers have drawn attention to the spatial structure of star formation 
in tidally induced starburst galaxies (Meurer 1995): 
stars form in clusters of 10$^5$ to 10$^7$\,\Msol, themselves 
associated in super-clusters (SCs). The survival of some of these
clusters for more than 10$^8$\,yrs is being investigated in relation
with globular cluster formation theories. On the basis of broad band
colours, Miller et al. (1997) attribute ages between 0.5 and 1\,Gyr to the 
most massive clusters of the post-starburst galaxy NGC 7252. Our
simulations indicate that AGB features should be significant in
these objects, unless the prescription adopted for stellar evolution through 
the AGB  is not relevant for starburst environments.
Kroupa (1998) suggests a slightly different evolution for starburst
clusters, possibly leading to dwarf elliptical galaxies. According to
his dynamical arguments, about half of the clusters in an SC merge within 
$\sim 10^8$\,yrs, forming bound objects of 10$^6$ to several 10$^8$\,\Msol 
with radii up to a few 100\,pc. Assuming a surface density of  
the order of 100\,\Msol/pc$^2$ for the underlying galaxy population
(twice the solar neighbourhood value), the AGB signatures are again
expected to show through in most of these cases.
The more extended post-burst objects predicted by 
Kroupa may have been overlooked until now in studies based on optical
surface brightness such as the NGC\,7252 survey of Miller et al. (1997).
The near-IR signatures of AGB stars could help us determine and understand
the fate of the SCs and be used more generally in  studies of the 
propagation of star formation in galaxies. Selected dwarf galaxies
(e.g. Tol 1924-416, \"Ostlin et al. 1998) fall in this category of studies.

Going one step ``further out", Canalizo \& Stockton (1997)
argue that the optical spectrum of the 
companion galaxy of QSO PG 1700+518 ($z=0.29$) is dominated by the 
emission of $\sim 10^8$\,yr old stars. 
Objects of this type will extend our stellar evolution laboratories
to the extreme environments of the vicinity of active galactic nuclei.

As we have shown, the separation between old populations dominated by 
red giants and
younger, post-starburst populations dominated by AGB stars, based on
the molecular bands discussed in this paper, opens new perspectives in the
exploitation of the near-IR galaxy spectra now becoming available. Recent
mid-IR simulations including the circumstellar emission of late AGB
stars (Bressan et al. 1998) suggest that combined mid- and
near-IR studies will be particularly helpful in solving degeneracies
between the effects of stellar ages, extinction and metallicity.
We note however that at the current state of knowledge the new
observations, in particular those of massive post-starburst clusters,
should also be seen as essential constraints on the AGB evolution models,
used as input for population synthesis predictions.


\begin{acknowledgements}
We address many thanks to P.\,Wood, R. Alvarez, M.\,Groenewegen and
B.\,Rocca-Volmerange for
their contributions to the progress of this work. 
\end{acknowledgements}

\end{document}